# Effect of induced shape anisotropy on magnetic properties of ferromagnetic cobalt nanocubes

D. Srikala[1*], V. N. Singh[2], A. Banerjee[3] and B. R. Mehta[2]

[1] School of Physical Sciences, Jawaharlal Nehru University, New Delhi 110067, India

[2] Thin Film Laboratory, Department of Physics, Indian Institute of Technology Delhi, New Delhi 110016, India

[3] UGC-DAE Consortium for Scientific Research, University Campus, Khandwa Road, Indore 452017, India

Corresponding author mail id: srikala.d@gmail.com

**Abstract.** We report on the synthesis of ferromagnetic cobalt nanocubes of various sizes using thermal pyrolysis method and the effect of shape anisotropy on the static and dynamic magnetic properties were studied. Shape anisotropy of approximately 10 % was introduced in nanocubes by making nanodiscs using a linear chain amine surfactant during synthesis process. It has been observed that, ferromagnetism persisted above room temperature and a sharp drop in magnetic moment at low temperatures in zero-field cooled magnetization may be associated with the spin disorder due to the effective anisotropy present in the system. Dynamic magnetic properties were studied using RF transverse susceptibility measurements at different temperatures and the singularities due to anisotropy fields were probed at low temperatures. Symmetrically located broad peaks are observed in the frozen state at the effective anisotropy fields and the peak structure is strongly affected by shape anisotropy and temperature. Irrespective of size the shape anisotropy gave rise to higher coercive fields and larger transverse susceptibility ratio at all temperatures. The role of shape anisotropy and the size of the particles on the observed magnetic behaviour were discussed.

# 1. Introduction

Diverse technological applications of magnetic nanoparticles in recording tape, flexible disk recording media, permanent magnets, microwave oscillators as well as bio medical materials and catalysts provides an impetus for extensive research in nanometre scale magnets.[1-3] Most of these applications rely on the stability of magnetic ordering with time. In nanometre scale magnets, the thermal fluctuations randomize the magnetic moment by overcoming the anisotropy energy barrier leading to unstable state of superparamagnetism. Once the anisotropy energy becomes comparable to a few KT, the coercivity and hysteresis falls to zero. Thus the demand for further miniaturization of nanomaterials comes into conflict due to the reduction of anisotropy energy per particle. Indeed miniaturization is required to achieve higher performance and lower cost in solid-state electronics, where spin transfer has potential to replace field-driven switching in magnetic memory and enables higher storage capacity.[4] Anisotropy is important in nanosystems because it helps in reducing the magnitude of thermal fluctuations (i.e., superparamagnetic state). This anisotropy in magnetic nanostructured materials depends not only on the band structure of the parent material, but also on the shape of the nanoparticles.[5] The most common anisotropies in nanostructured materials are crystalline and shape anisotropy. The unidirectional magnetocrystalline anisotropy at the interface of ferromagnet and antiferromagnet stabilizes the system to high temperatures.[6-9] High anisotropy materials with coercivity proportional to anisotropy constant are attractive candidates for high-coercivity applications and shape anisotropy is predicted to produce the largest coercive forces.[5] The shape anisotropy is a phenomenon associated with the zeroth spatial order of the magnetization field and it has great advantage in aligning the magnetic axis of the particles with respect to the surface during self-assembly into high quality 2D and 3D arrays.[10] In magnetic material, the need for elongated structures (shape anisotropy) is paramount and even small deviations from spherical shapes can significantly alter the coericity.[11] It is known that not only size but the shape of the metal nanoparticles also influences the energy spectrum, surface charge density, colloidal stability and magnetic properties of the materials. In this paper, we study the effect of shape anisotropy in cobalt nanocubes using static and dynamic magnetic properties at various temperatures. The shape anisotropy was introduced in the form of nanodiscs by using a linear chain amine surfactant in the growth process. The variations in particle size and shape induced predominant changes in the magnetic behaviour of nanocubes. We observed the persistence of ferromagnetism above room temperature. Larger coercive field and RF transverse susceptibility ratio's were measured and have been correlated with the shape anisotropy present in nanocubes.

## 2. Experiments

Cobalt nanocubes were prepared in inert atmosphere using a standard pyrolysis procedure.[12, 13] The synthesis involves thermal decomposition of $Co_2(CO)_8$ in a solvent in the presence of surfactants. Surfactant solution was heated to desired temperature depending on the objective of the experiment and the precursor solution was rapidly injected into it. For the synthesis of various cubic Co nanoparticles the following chemicals were used: Di-cobalt octa-carbonyl $Co_2(CO)_8$ containing 1-5% hexane as a stabilizer, o-dichlorobenzene (DCB, 99 %), trioctylphosphine oxide (TOPO, 99%), octadecylamine (ODA, 97%), O,O′-bis(2-aminopropyl)polypropyleneglycol (Jeffamine D-400), and octanoic acid (99.5 %).

For the synthesis of nanocubes the DCB was first dried with anhydrous $CaCl_2$ powder and stored under vacuum. A solution containing the surfactants, 0.54 mL O,O'-bis(2-aminopropyl) polypropyleneglycol, 0.09 mL octanoic acid, and 0.12 g TOPO in 13 mL DCB, was degassed for 1 h and then heated to reflux temperature (~186 $^0$C) in the presence of high purity argon gas. Then, 0.62 g of $Co_2(CO)_8$ were dissolved in 4 mL DCB and rapidly injected in to the reflux bath. The solution was allowed to reflux for 3 min and cooled down in liquid nitrogen. Same process was followed for 1 min reaction time to make smaller nanocubes. Synthesis of nanocubes along with discs was carried out by first degassing 0.6 g ODA and 0.2 g TOPO with argon gas in three neck flask for 20 min. 15 mL of o-DCB was introduced into the flask and heated to ~ 182$^0$C. 0.54 g of $Co_2(CO)_8$ diluted in 3 mL of o-DCB is rapidly injected into the refluxed bath and the reaction was carried out for 3 min.

After the reaction, the nanoparticles were extracted from the colloidal solution, by adding an equal volume of ethanol and hexane, centrifuged for 20 min at a speed of 2100 rpm/min and vacuum dried. Size and shape of Co nanoparticles were studied by transmission electron microscopy (TEM) using a technaiG$^{20}$ (200 KV) microscope and the elemental composition study was estimated using energy dispersive x-ray analysis attached to it. The crystallographic phases were investigated using Rigaku X-ray diffractometer (model Giegerflex-D/max-RB-RU200). Magnetization measurements were carried out on a Quantum Design physical property measurement system in the temperature range from 2 to 350 K and in applied magnetic fields up to 7 Tesla. The transverse susceptibility measurements were done using a tunnel diode oscillator technique to study the dynamic magnetic properties.

## 3. Results and discussion

Cobalt nanocubes of different sizes were prepared at different reaction conditions and are shown in figures 1 and 2. Shape anisotropy in the form of discs was introduced in cubic particles to study their effect on magnetic properties by decreasing the size of the nanocubes. Synthesis of pure nanocubes was carried out by keeping the surfactant to precursor ratio constant at reaction times of 1 min and 3 min respectively. 45 nm sized nanocubes (named as $NC_1$) were formed for a reaction time of 3 min and 30 nm ($NC_2$) sized cubes for 1min reaction time as shown in Figure 1 (a) and Figure 1 (b). If the reaction is quenched in the earliest moments (few seconds) after the growth begins, then smaller sized cubes can be obtained; however, it is difficult to quench the reaction so quickly. Inset of Figure1 (b) shows the electron diffraction rings indicating the formation of ε-Co structure in $NC_2$. Figure 1 (c), shows an x-ray diffraction pattern of $NC_1$ exhibiting a combination of fcc - *hcp* Co structure. The observed peaks in the x-ray diffraction: (002), and (101) correspond to *hcp* Co and (200) correspond to fcc Co. The particles size estimated using Scherrer's formula matches well with the size observed using TEM.

To introduce shape anisotropy in nanocubes, two surfactants O,O′-bis(2-aminopropyl)polypropyleneglycol, and octanoic acid used for preparing pure nanocubes were replaced by an linear amine surfactant octadecylamine (ODA). The surfactants ODA and TOPO yielded combination of cube and disc shaped nanoparticles as shown in Figure 2. The average size of discs formed were 4 nm × 19.3 nm and cubes of edge length 19.6 nm ($NC_3$). Yield of nanodiscs is around 10 % of nanocubes. It is noticed that, $-NH_2$ group of amine surfactant was responsible for formation of disc shaped cobalt nanoparticles. It is shown by *V. F. Puntes* et al., that disc yield can be further enhanced by using larger amounts of surfactants or when amine is co-injected with the cobalt precursor.[13] Most of the cobalt nanodiscs are self- assembled into ribbon like structure, in a way that the larger discs are at the centre of the ribbon when dispersed on TEM grid as shown in inset of Figure 2 (a). Figure 2 (b) shows the electron diffraction pattern as identified by a crystal structure with the symmetry found in the β phase of elemental manganese with 20 atoms present in a cube of 6.09 Å, first named by Bawendi[14] as ε-cobalt and observed by other groups as well.[15, 16] The observed rings corresponds to (211), (220), (332), (530), and (610) crystal planes of ε-Co. Figure 2 (c) displays the high resolution TEM image with different crystal planes. As the time evolves the size and shape of nanocubes appears to be determined by the type of surfactant and its concentration, which is consistent with observations in many other metal nanoparticles.[17]

A series of zero-field-cooled and field-cooled scans of all the three samples in the presence of 100 Oe magnetic field are shown in Figure 3, in which magnetization is measured as a function of temperature. The temperature dependence of the magnetization is consistent with a scenario in which the nanocubes are superparamagnetic above ~ 350 K, with the moments blocked at lower temperatures. The sharp drop in ZFC magnetization below ~ 13 K is due to the thermal blocking of particles moment in random directions. Above this temperature there is a gradual increase in the magnetization up to the measured temperature 350 K which suggests a progressive rotation of the magnetization of the thermally blocked particles towards the field direction. None of the FC and ZFC curves show any characteristic sharp change in magnetization associated with the well established ferromagnetic to superparamagnetic transition. The significant feature observed in the ZFC magnetization data of sample $NC_2$ as shown in Figure 3 (b), is a peak at 23 K and the corresponding FC curve showing sharp increase in magnetic moment which saturates at low temperatures. The increase in FC magnetic moment can be understood as the spins frozen in the field direction and the sharp drop in ZFC magnetization below the peak at lower temperatures is expected when the surface spins freeze in random orientations resulting in a spin-glass like behaviour which may have arisen out of disorder and frustration. Generally, the ZFC magnetization shows a peak for both superparamagnetic and for spin glass behaviour and the FC magnetization saturates below the spin glass peak temperature.[18] The magnetization of the samples increases with the increase in size of the particles irrespective of the shape anisotropy present in $NC_3$ up to 300 K. A sudden rise in the ZFC magnetization of $NC_3$ above 320 K as shown in Figure 3 (c) indicates the relaxation phenomena present in the system. The alignment of spins in the presence of small applied fields in the elongated particles i.e., in nanodiscs gave rise to a high moment up to the measured temperature of 350 K, whereas this feature is not observed in pure cubic samples.

DC magnetization measurements on Co nanocubes revealed a very small coercive field at 10 K and as well as at 300 K as shown in Figure 4. Coercive field $H_C$ of 230 Oe is observed in 45 nm sized ($NC_1$) cubic particles and maximum coercivity of 318 Oe is observed in $NC_3$ at 10 K as shown in inset of Figure 4 (c). $H_C$ decreased only very little, by about 5 % in 30 nm sized cubes. This deviation in $H_C$ as compared to the cubic shape particles can be coupled to the shape anisotropy present in sample and it refers to the presence of preferred magnetization direction within the material which is ultimately responsible in determining the magnetic properties of the sample. The observed variation in $H_C$ though the size of the cubes is smaller is because of coherent rotation of spins in the entire stack of nanodiscs. The coercive field $H_C$ at 10 K and at 300 K is almost similar which is consistent with the scenario in which the nanocubes are ferromagnetic above ~ 300 K. The saturation magnetization decreases as the size of the cubes decreased because of increase in nanoparticle surface-to-volume ratio and absorption of ligands at the surface of the nanocrystals which affects the electronic and magnetic structures at the surface and thereby impacts on the magnetic properties of the system.[19, 20] The saturation magnetization $M_S$ of $NC_1$ takes on a value of about 121 emu/g at 10 K which is only 75 % of bulk $M_S$ of Co (~ 162 emu/g[21]). 44.9 emu/g and 49.4 emu/g were observed in $NC_2$ and $NC_3$. Using the total saturation magnetization at 4 T the magnetic moment per particle at 10 K was calculated from $\chi = M_S\mu/ 3k_BT$, where $\chi$ is the susceptibility, $\mu$ is the magnetic moment per particle, and $k_B$ is Boltzmann's constant. Highest magnetic moment per particle of 186 $\mu_B$ is obtained in $NC_1$. The effective magnetic moment per particle increases with the increase in the size of nanocubes due to canting of the surface spins.[19, 21] The measured ZFC MH loops at 10 K were symmetric along the field axis. In order to further test the idea of exchange bias effect we have cooled down the samples from 350 K to 10 K in an applied field of 2 T and the field is scanned from -7 T to +7 T as shown in Figure 5. As expected the field cooling procedure resulted in no exchange bias effect confirming the stable and un-oxidized Co nanocubes. Symmetric ZFC and FC hysteresis loops are shown in inset of Figure 5.

Anisotropy is the one which determines the magnitude of the energy barrier $\Delta E = KV$, where K is the effective anisotropy energy and V is the particle volume. The effective anisotropy energy K is estimated using $K = 30K_B T_B/V$, where $K_B$ is the Boltzmann constant and $T_B$ is the blocking temperature taken at 350 K. As the particle size decreased, the anisotropy increased. 45 nm sized cubes have minimum anisotropy of $0.16 \times 10^5$ erg/Cm$^3$ and as the particles size decreased to 19.6 nm the anisotropy increased by almost 12 times. The possible source of enormous increase in anisotropy in NC$_3$ is surface anisotropy which is because of finite size effect and shape anisotropy. Since it is known that total anisotropy energy barrier depends on volume and surface anisotropy energy densities[23], so for a given volume of a particle the surface area is more for elongated shaped particles. Hence the major contribution from surface to the effective anisotropy and an increase in coercivity is also expected in elongated particles. Stress anisotropy which results from internal or external stresses during synthesis may also have some contribution to the total value of anisotropy.

In order to study the effect of shape anisotropy on magnetic properties in finer details the dynamic susceptibility measurements were performed using a transverse susceptibility method. This measurement provides sensitive and unique way to understand the influence of the relaxation, interactions and other phenomena that govern the dynamic magnetic properties in magnetic nanostructures.[24] It is probed using a tunnel diode oscillator technique operating at radio frequencies. The sample in powder form is packed in a nonmagnetic tube that fits into the core of inductive copper coil. This is inserted into the sample space of a commercial cryogen free system using rf coaxial cables perpendicular to the varying external dc field and this arrangement sets up the transverse geometry. The shift in resonant frequency is measured as the external dc field is varied at different temperatures. Since the sample is kept in an rf coil that is a part of self-resonant circuit, the shift in the frequency with varying dc field gives a direct measure of the change in inductance of the coil, determined by the change in transverse permeability $\mu_T$ of the sample.[25] The transverse susceptibility and the real part of the complex permeability are related via the expression $\chi_T = \mu_T - 1$. Thus knowing the precise geometrical parameters the absolute value of the transverse susceptibility can be derived. Here $\chi_T$ is the percentage change in the resonant frequency, which is a figure of merit, and is used to study the variations in magnetic properties at different temperatures.

The transverse susceptibility ratio ($\Delta\chi_T/\chi_T$ (%)), measured in dc field from positive to negative field ($\pm$ 2 T) at fixed temperatures for all the samples in the temperature range 6 – 200 K are shown in Figure 6. The observed plots represent the unipolar field scan from positive to negative fields. The saturation field is taken at 2 T for all the samples. A prominent feature that can be distinguished clearly is at T = 6.5 K where the curves show two peaks; a behaviour which occurs at the effective anisotropy fields of these materials. These low field peaks differ in their heights showing asymmetry in a unipolar scan and are considerably broadened because of the distribution in particle size and anisotropy fields of the collection of particles. In $NC_2$ and $NC_3$ the lower field peak in the negative field axis has larger height than the positive peak as shown in Figure 6 (b) and Figure 6 (c), whereas in $NC_1$ the larger peak was located at positive field (Figure 6 (a)). No sharp asymmetry in peak heights is observed in larger sized nanocubes ($NC_1$). When the field is swept from positive saturation to zero most of the spins contribute to the overall susceptibility at the anisotropy field. As the field passes through zero to negative the spins begin to align in the negative direction of the field and due to unequal switching of the magnetic moments between the equilibrium positions an asymmetry is observed in peak heights. The observed peaks at 6.5 K represent singularities due to the anisotropy fields. In $NC_2$ at 6.5 K the negative saturation is not reached as shown in Figure 6 (b), because of random orientation of spins being blocked as observed in the dc magnetization with a fall in magnetic moment below the peak temperature. At low temperature i.e., in the frozen state almost all the particles of the system have a relaxation time much greater than the experimental time. Thus the two peak behaviour determines the irreversible nature of the $\chi_T$. Theory of Aharoni based on the Stoner-Wohlfarth model predicts the existence of three singularities in a unipolar field scan from positive to negative saturation, two at the anisotropy fields ($\pm H_K$) and one at the switching field.[26] But this theory is applicable for non interacting single-domain magnetic particles and does not take into consideration inter-particle interactions, particle size and shape dispersion, and relaxation effects. However, most of the experimental studies usually report two symmetric peaks located at the anisotropy fields and the third peak is often broadened and merged with one of these peaks due to distribution in particle size and these peaks are highly sensitive to surface magnetic behaviour, particle size and shape dispersion, magnetocrystalline nature, and sample temperature.[27] An increase in the zero field value of $\chi_T$ is noticed as the temperature increased up to 200 K and no signature of the transition from ferromagnetic to superparamagnetic nature is found, which is consistent with the ferromagnetic nature observed in our dc magnetization measurements. Thus $\chi_T$ confirms the existence of ferromagnetism up to high temperatures. As the temperature increases, the double peak structure at the effective anisotropy fields becomes less pronounced and merges into a single peak at much lower field values corresponding to the switching fields where thermal energy dominates over anisotropic energy.

In $NC_3$, though the size of the cubes and discs are smaller, there is not only prominent asymmetry in peak heights but also have a large $\Delta\chi_T/\chi_T$ magnitude, which increases drastically when the temperature is varied from 6.5 K to 40.5 K as shown in Figure 6 (c). The interaction between spins leading to the effective anisotropy has a significant contribution from shape anisotropy and the surface magnetic spin nature. We emphasize the point that this characteristic of cube and disc shape nanoparticles, confirms that the shape anisotropy determined the transverse susceptibility ratio. The magnetic field is cycled with the sample held at all the five temperatures i.e., 6.5 K, 40.5 K, 101 K, 151 K, and 200 K and the field is scanned from zero to + 2 T and + 2 T to – 2 T. The peak shift and the hysteresis are clearly visible in the transverse susceptibility ratio plot at T = 6.5 K, 40.5 K, and 200 K as shown in Figure 7. As the field is varied an asymmetry in the peak heights and position is observed in all the samples which clearly indicates the irreversible behaviour of magnetic moments.

In conclusion cobalt nanocubes were synthesized by thermal decomposition of dicobalt octacarbonyl in the presence of various surfactants. Static and dynamic magnetization measurements revealed strong size and shape dependent properties along with persistence of ferromagnetism at high temperatures. Singularities due to anisotropy fields were observed at the lowest temperature in different magnitudes and as the thermal energy dominated the peak structure followed the dc magnetization. RF transverse susceptibility ratio and coercivity were found to be maximum in case of shape anisotropy. The fact that one sees such strong variations in these parameters, suggests that the shape anisotropy plays an essential role in determining the magnetic properties of nanoparticles.

**Acknowledgements:** D. Srikala acknowledges UGC for financial support. DST, Government of India is acknowledged for funding the VSM at CSR, Indore and High field magnet facility at JNU, New Delhi.

**Figure Captions:**

**Fig. 1.** TEM micrograph of (a) nanocubes of average length 44.5 nm for a reaction time of 3 min (NC$_1$), (b) 30 nm for a reaction time of 1 min (NC$_2$); inset shows the crystalline nature, and (c) X-ray diffractogram of Co nanocube powder sample NC$_1$. Bragg peaks labelled as (002) and (101) comply with *hcp* Co, whereas * marked peak labelled (200) comply with fcc Co.

**Fig. 2.** (a) TEM micrograph of cobalt nanocubes and nanodics (NC$_3$). Inset shows the self assembly of discs, (b) Electron diffraction micrograph exhibiting crystalline nature corresponding to the ε-Co structure, and (C) HRTEM image showing different crystal planes.

**Fig. 3.** (Color online) Temperature dependence of the zero-field-cooled and field cooled magnetization in 100 Oe magnetic field of (a) nanocubes of 44.5 nm (b) nanocubes of 30 nm, (c) combination of nanocubes and nanodiscs, exhibiting ferromagnetic behaviour up to 350 K.

**Fig. 4.** (Color online) Hysteresis loops of (a) NC$_1$, (b) NC$_2$, and (c) NC$_3$ at 300 K and 10 K. The insets show the magnetization curves exhibiting similar coercivity at both the temperatures.

**Fig. 5.** (Color online) Zero field cooled and 2 T field cooled M - H loops of nanocubes and discs exhibiting no exchange bias effect. Inset shows the symmetric hysteresis loops.

**Fig. 6.** (Color online) RF transverse susceptibility ratio of (a) NC$_1$, (b) NC$_2$, and (c) NC$_3$ at T = 6.5 K (■), 40.5 K (●), 101 K (♦), 151 K (◀), and 200 K (∗).

**Fig. 7.** (Color online) Peak structure and the hysteresis behaviour of transverse susceptibility in field scans at T = 6.5 K, 40.5 K and 200 K.

**Fig. 1.**

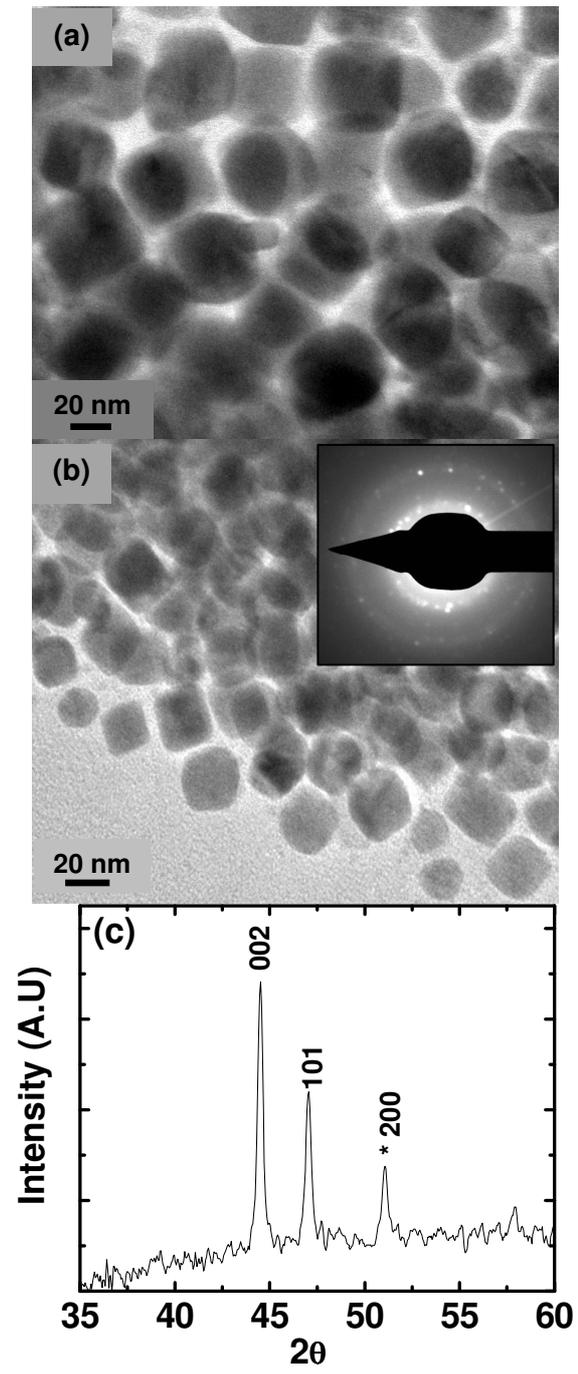

**Fig. 2.**

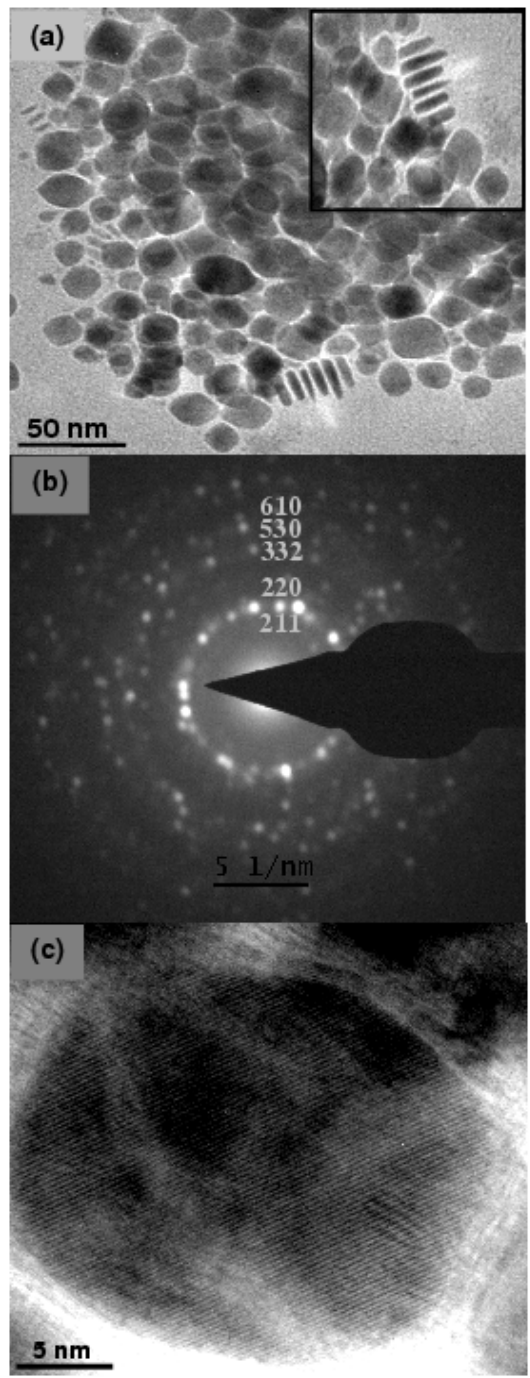

**Fig. 3.**

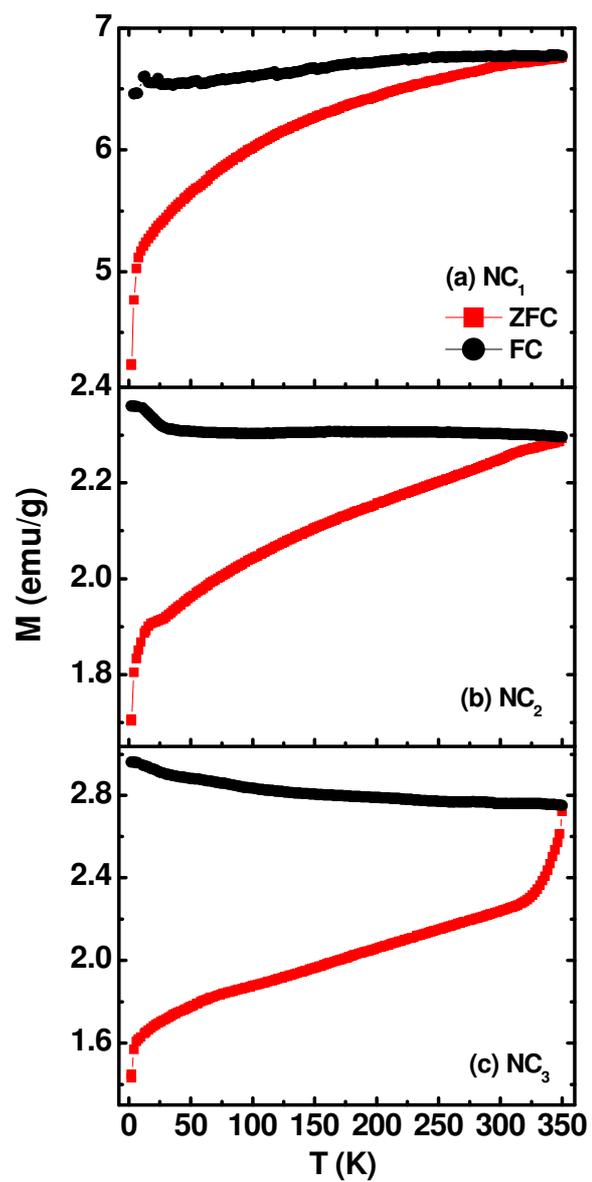

**Fig. 4.**

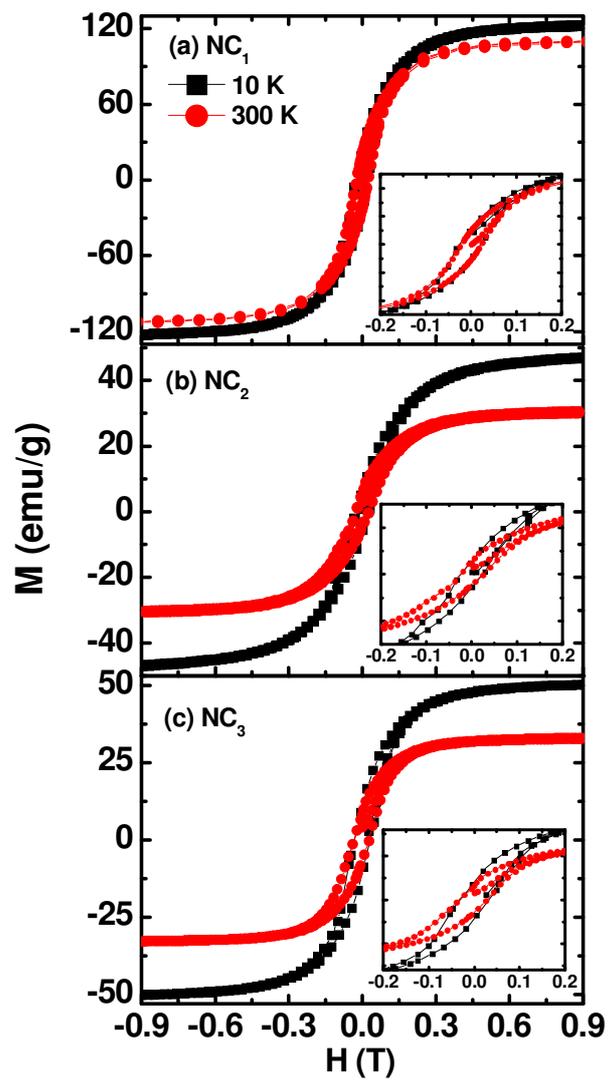

**Fig. 5.**

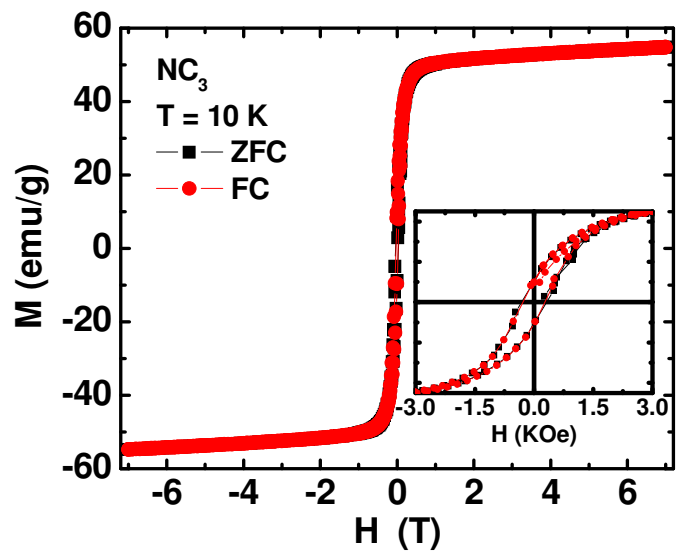

**Fig. 6.**

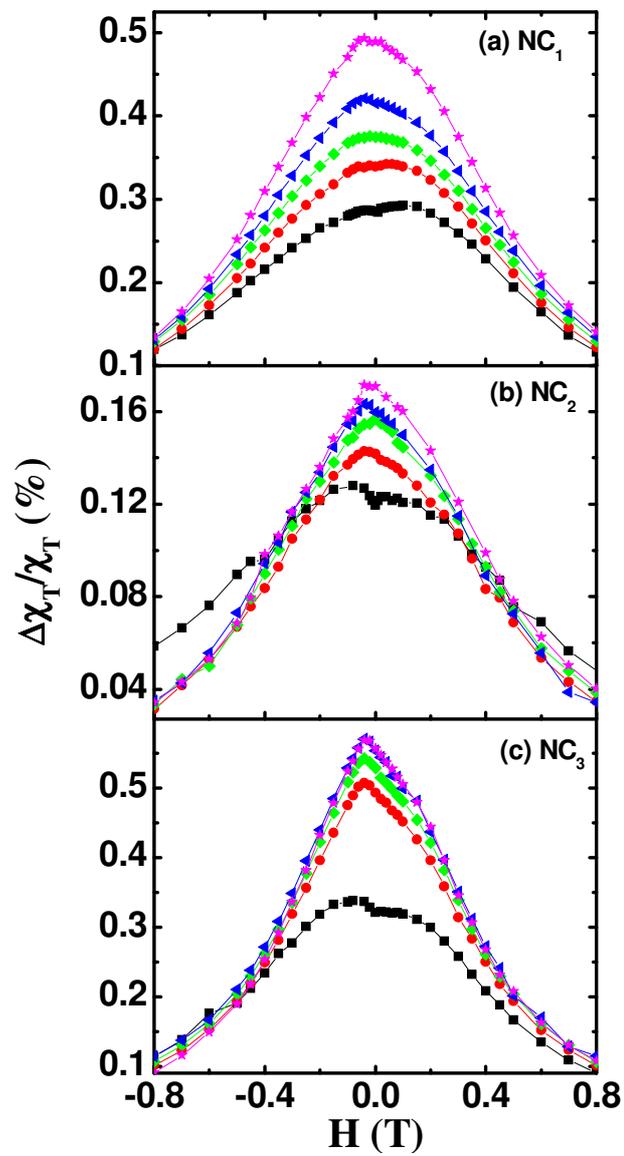

**Fig. 7.**

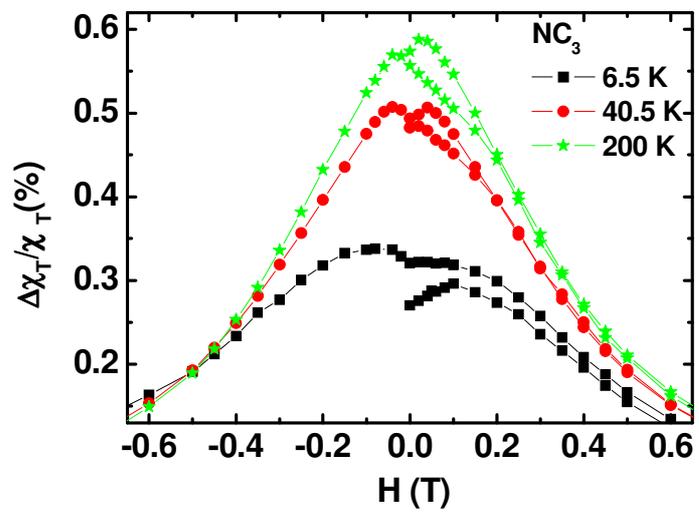